\documentclass[twocolumn,showkeys,showpacs,preprintnumbers,prd,superscriptaddress,nofootinbib]{revtex4-1}
\usepackage{graphicx}
\usepackage{dcolumn}
\usepackage{bm}
\usepackage{hyperref}

\usepackage{newunicodechar}
\usepackage{amsmath}
\newunicodechar{−}{\ensuremath{-}}

\usepackage[dvipsnames]{xcolor}
\definecolor{darkblue}{rgb}{0.0, 0.0, 0.55}
\definecolor{darkred}{rgb}{0.55, 0.0, 0.0}
\usepackage{cleveref}
\hypersetup{
    colorlinks=true,
    linkcolor=darkblue,
    citecolor=darkblue,
    urlcolor=darkblue}

\begin{document}

%\preprint{APS/123-QED}

\title{Probing dynamical embeddings in a five-dimensional spacetime in light of DESI BAO}

\author{Abra\~{a}o J. S. Capistrano}\email{capistrano@ufpr.br}
\affiliation{Universidade Federal do Paran\'{a}, Departmento de Engenharia e Exatas, Rua Pioneiro, 2153, Palotina, 85950-000, Paraná/PR, Brasil\\
Federal University of Latin American Integration (UNILA), Applied physics graduation program, Avenida Tarqu\'{i}nio Joslin dos Santos, 1000-Polo Universit\'{a}rio, Foz do Igua\c{c}u, 85867-670, Paran\'{a}/PR,Brasil}

\author{Emanuelly Silva}
\email{emanuelly.santos@ufrgs.br}
\affiliation{Instituto de F\'{i}sica, Universidade Federal do Rio Grande do Sul, 91501-970 Porto Alegre RS, Brazil}

\author{Rafael C. Nunes}
\email{rafadcnunes@gmail.com}
\affiliation{Instituto de F\'{i}sica, Universidade Federal do Rio Grande do Sul, 91501-970 Porto Alegre RS, Brazil}
\affiliation{Divisão de Astrofísica, Instituto Nacional de Pesquisas Espaciais, Avenida dos Astronautas 1758, São José dos Campos, 12227-010, São Paulo, Brazil}

\author{Orlando Luongo}
\email{orlando.luongo@unicam.it}
\affiliation{Department of Nanoscale Science and Engineering, University at Albany-SUNY, Albany, New York 12222, USA.}
\affiliation{Universit\`a di Camerino, Divisione di Fisica, Via Madonna delle carceri 9, 62032 Camerino, Italy.}
\affiliation{INFN, Sezione di Perugia, Perugia, 06123, Italy.}
\affiliation{INAF - Osservatorio Astronomico di Brera, Milano, Italy.}
\affiliation{Al-Farabi Kazakh National University, Al-Farabi av. 71, 050040 Almaty, Kazakhstan.}

\date{\today}

\begin{abstract}
We here investigate the observational viability of Nash gravity as an alternative to the standard $\Lambda$CDM cosmology. Based on Nash's embedding theorem, the model introduces orthogonal perturbations via variations in the extrinsic curvature, generating scalar-type metric perturbations directly from geometry, without the need to introduce additional fields. We confront the model with current observational data, including Cosmic Microwave Background (CMB) measurements from Planck, Baryon Acoustic Oscillations (BAO) from DESI DR2, and recent Type Ia supernova (SN Ia) compilations. Our analysis shows that Nash gravity provides a good fit to the data, yielding a slightly higher value for the Hubble constant, $H_0 = 69.32 \pm 0.72$ km/s/Mpc, compared to the $\Lambda$CDM model, thus offering a potential alleviation of the $H_0$ tension. Furthermore, the model naturally predicts a suppressed growth of structure, with $S_8 \approx 0.76$ across various joint analyses, potentially alleviating the so-called $S_8$ tension, assuming that this discrepancy is not solely due to systematic effects in other independent measurements. In some cases, Nash gravity achieves a better fit to the data than the $\Lambda$CDM paradigm at the $2\sigma$ level.
\end{abstract}

\maketitle

\section{Introduction}

Several modifications of General Relativity (GR) have been introduced and thoroughly explored as attempts to address longstanding observational issues in cosmology and astrophysics (see, e.g.,~\cite{Clifton:2011jh,Frusciante:2019xia,Ishak:2018his} for detailed reviews). In particular, modified gravity  theories provide a broader theoretical framework that extends the standard $\Lambda$CDM scenario, offering alternative explanations for the Universe's late-time accelerated expansion without necessarily invoking a cosmological constant $\Lambda$. Despite the fact that many of these models yield excellent agreement with current observations, they often exhibit theoretical and observational degeneracies. That is, distinct models can produce nearly equivalent observational predictions, making it challenging to differentiate between them using existing cosmological data.

Nevertheless, the growing accuracy of cosmological measurements has begun to put pressure on the internal consistency of the $\Lambda$CDM paradigm, giving rise to several well-known tensions among independent datasets~\cite{Perivolaropoulos:2021jda,Akarsu:2024qiq,CosmoVerse:2025txj,DiValentino:2021izs,Vagnozzi:2023nrq}. Notably, discrepancies in the inferred values of the Hubble constant, \( H_0 \), and the matter clustering amplitude, \( S_8 \), have drawn considerable attention. These tensions may point toward new physics beyond $\Lambda$CDM, and potentially beyond standard GR. In this context, various extensions of GR have gained prominence as promising alternatives to address such discrepancies (see, for instance,~\cite{DiValentino:2024wgi,DeFelice:2020cpt,deAraujo:2021cnd,Moshafi:2020rkq,Pogosian:2021mcs,Braglia:2020auw,Petronikolou:2021shp,Briffa:2023ozo,Escamilla:2024xmz,Aljaf:2022fbk,Lopez-Hernandez:2024osv,Odintsov:2020qzd,Benevento:2022cql,DAgostino:2020dhv,Boiza:2025xpn,Toda:2024fgv,Bouche:2022qcv} and~\cite{CosmoVerse:2025txj} for general discussions). The observational imprints associated with these tensions could serve as powerful probes of deviations from GR and may ultimately provide important clues about the nature of gravity on cosmological scales. Possible signatures of modified gravity in cosmological data have previously played an important role in the literature (e.g., \cite{Bonvin:2025qeo,Ishak:2024jhs,Plaza:2025ryz,Rodriguez-Meza:2023rga,Aviles:2024zlw,Yan:2024jwz,Scherer:2025esj,Wang:2024uvw,Castello:2024jmq,Pan:2025psn,Specogna:2024euz,Akarsu:2024eoo,Aoki:2020oqc,Wolf:2025jed,Yang:2025mws,Odintsov:2025kyw,Escamilla-Rivera:2024sae,Kumar:2025mzo,BarrosoVarela:2024ozs,Leizerovich:2021ksf,Tiwari:2024gzo}). 

This paper is not M-theory or string-related models but instead focuses on the dynamics of space-time embeddings. The Randall-Sundrum (RS) model~\cite{Randall1999}, a brane-world theory derived from M-theory, utilizes a five-dimensional anti-de Sitter (\(\text{AdS}_5\)) space, by the prospects of the AdS/CFT correspondence between the superconformal Yang-Mills theory in four dimensions and the anti-deSitter gravity in five dimensions. In the Randall-Sundrum model II~\cite{Randall1999b}, a \(\mathcal{Z}_2\) symmetry is applied across a background brane-world, where the Israel-Lanczos condition~\cite{Lanczos1924,Israel1966} replaces the extrinsic curvature by the terms of the energy-momentum tensor of its confined sources. However, this condition is not universally applied in all brane-world models. Some alternative approaches, such as those by Arkani-Hamed, Dvali, and Dimopoulos~\cite{ArkaniHamed1998}, as well as the Dvali-Gabadadze-Porrati model~\cite{Dvali2000}, use different junction conditions, leading to distinct physical consequences. Other brane-world models do not require specific junction conditions, showing that the Israel condition is not unique and may not be fundamental to brane-world cosmology~\cite{Battye2001,Davis2003,Yamauchi2007,Jalalzadeh2009,Heydari-Fard2010,Rashki2012}. Additionally, the Israel condition, being particular to five-dimensional bulk spaces, does not extend naturally to higher-dimensional embeddings, such as those involving a Schwarzschild solution~\cite{Kasner1921}. In these cases, the extrinsic curvature cannot be uniquely determined by a single confined energy-momentum tensor, limiting the applicability of the condition. 
 
In a covariant and model-independent approach, Nash's embedding theorem provides a foundation for the embedding of Riemannian geometries, demonstrating that any unperturbed metric can be obtained through a continuous sequence of small perturbations~\cite{Nash1956}. Later extended to pseudo-Riemannian manifolds~\cite{Greene1970}, we use Nash's embedding theorem as tool to gravitational perturbation theory, which the geometric flow and extrinsic curvature are regarded as dynamical variables, distinguishing it from conventional brane-world scenarios.  

In this paper, we introduce a model that incorporates an additional curvature component into standard Einstein gravity through the mechanism of local dynamical embeddings. Unlike conventional extra-dimensional braneworld scenarios~\cite{ArkaniHamed1998,Randall1999,Randall1999b,Dvali2000}, our framework explicitly includes the dynamics of the extrinsic curvature as part of the gravitational description. The primary aim of this work is to investigate how a dynamically evolving embedding can influence both the background spacetime geometry and the behavior of scalar perturbations at the linear level. 

The present work revises and simplifies this framework, unifying background and perturbative dynamics. We extend the analysis of previous papers ~\cite{MAIA20029,GDE,Maia_2007,gde2,Capistrano2021,capistrano2022,Capistrano2024} and new theoretical advances can be summarized as follows:

\begin{itemize}
\item Reformulation of the 5D embedding dynamics directly in terms of the extrinsic curvature scalar $k_{\mu\nu}k^{\mu\nu}$ and its dependence on the scale factor $a(t)$, providing a simpler and more transparent dynamical framework that links geometry and cosmology;
\item Introduction of a new dimensionless coupling $\beta_0$, which captures the residual effects of extrinsic curvature perturbations and clarifies deviations from GR growth dynamics;
\item Derivation of the full set of linear perturbation equations for Nash Gravity (Eqs.~\ref{tensorcompo00kspace}--\ref{muequation}), incorporating the coupled evolution of the $\Phi$ and $\Psi$ potentials, a geometric modification of the Poisson equation via the $\beta_0$ parameter, and the resulting scale-independent effective gravitational constant $G_\text{eff}(a)$;
\item Indication that perturbations remain ghost-free, as the extrinsic curvature enters quadratically and satisfies the Codazzi conditions, explicitly verified through the Gauss–Codazzi constraints.
\end{itemize}

We then compare this new model with current cosmological observations, placing particular emphasis on the latest Baryon Acoustic Oscillation (BAO) measurements from the DESI (Dark Energy Spectroscopic Instrument) survey and their combination with other key cosmological datasets.
DESI is a cutting-edge spectroscopic survey specifically designed to chart the large-scale structure (LSS) of the Universe in three dimensions. It covers a vast area of the sky and spans a wide redshift range, providing an unprecedented volume of high-quality data. This dataset allows for the reconstruction of detailed three-dimensional maps of the cosmic web, supports high-precision determinations of the Universe's expansion history, and enables stringent tests of cosmological models that go beyond the standard $\Lambda$CDM framework~\cite{Carloni:2025jlk,Alfano:2025gie,Alfano:2024fzv,Alfano:2024jqn,Carloni:2024zpl,Luongo:2024fww}.

The paper is organized as follows. In Section~\ref{model}, we introduce the cosmological models considered in this work. Section~\ref{data} describes the observational data sets and the statistical methods employed in our analysis. In Section~\ref{results}, we present and discuss the main results, highlighting the implications of our parameter estimates for the model under investigation. Regarding notation, capital Latin indices run from 1 to 5 and denote quantities in the full embedding space. Lowercase Latin indices refer specifically to the single extra dimension considered. Greek indices span from 1 to 4 and are associated with the embedded four-dimensional space-time. A bar over a symbol denotes a background (unperturbed) quantity. Throughout this work, we adopt the Landau-Lifshitz metric signature convention, \( (+,-,-,-) \). Finally, Section~\ref{conclu} contains our concluding remarks and future perspectives.

\section{Nash embedding as a physical model}
\label{model}
In the following, we highlight the main aspects of embedding framework based on previous works~\cite{MAIA20029,GDE,Maia_2007,gde2,Capistrano2021,capistrano2022,Capistrano2024}. We consider a five-dimensional bulk $V_5$ endowed with an embedded four-dimensional geometry $V_4$~\cite{MAIA20029,GDE}. This higher-dimensional space is governed by the gravitational action $S$, defined as
\begin{equation}\label{eq:action}
S= -\frac{1}{2\kappa^2_5} \int \sqrt{|\mathcal{G}|}{\;}^5\mathcal{R}d^{5}x - \int \sqrt{|\mathcal{G}|}\mathcal{L}^{*}_{m}d^{5}x\;,
\end{equation}
where $\kappa^2_5$ represents the fundamental energy scale in the embedded space, $^5\mathcal{R}$ is the five-dimensional Ricci scalar, and $\mathcal{L}^{*}_{m}$ denotes the matter Lagrangian of fields confined to the four-dimensional space-time.

In this scenario, the main focus is on the appearance of the extrinsic curvature and its consequences for a physical model. The background extrinsic curvature $\bar{k}_{\mu\nu}$ is traditionally defined as~\cite{eisenhart1926riemannian}
\begin{equation}
\bar{k}_{\mu\nu} =  -\mathcal{X}^A_{,\mu}\;\bar{\eta}^B_{,\nu} \mathcal{G}_{AB}\;, \label{eq:extrinsic}
\end{equation}
which projects variations in the normal unitary vector $\bar{\eta}^B$ onto the tangent plane orthogonal to $V_4$. These variations govern the bending/stretching of $V_4$, yielding tangent components with $\bar{k}_{\mu\nu}$. Then,one defines an embedding coordinate $\mathcal{X}$ to a regular mapping $\mathcal{X}: V_4 \rightarrow V_5$. This must satisfy the set of the embedding non-perturbed equations
\begin{equation} \label{eq:X}
\mathcal{X}^A_{,\mu} \mathcal{X}^B_{,\nu}\mathcal{G}_{AB}=g_{\mu\nu},\;  \mathcal{X}^A_{,\mu}\bar{\eta}^B \mathcal{G}_{AB}=0,  \;  \bar{\eta}^A \bar{\eta}^B \mathcal{G}_{AB}=1\;,
\end{equation}
where $\mathcal{G}_{AB}$ represents the bulk metric components, which takes the form
\begin{eqnarray}
  \mathcal{G}_{AB} &=& \left(
          \begin{array}{cc}
            g_{\mu\nu} & 0 \\
            0 & 1 \\
          \end{array}
        \right)\;.
   \label{eq:metricbulk}
\end{eqnarray}

As well known, perturbations are essential in any physical model because they allow us to explore how small deviations from a background evolve making possible to detect the today's known or possible new observable quantities. This is particularly important for cosmology, where perturbations should explain the formation of LSS of the Universe and the anisotropies observed in the CMB~\cite{refplanck2018}. In the same spirit of Refs.\cite{MAIA20029,GDE}, we apply Nash's embedding theorem~\cite{Nash1956} to introduce orthogonal perturbations in the background metric $\bar{g}_{\mu\nu}$ as
\begin{equation}\label{eq:nashdeformation}
 \bar{k}_{\mu\nu}=-\frac{1}{2}\frac{\partial \bar{g}_{\mu\nu} }{\partial y}\;,
\end{equation}
where $y$ is the spatial coordinate orthogonal to the tangent plane. This provides a mechanism to account for the dynamics and stability of the embedded manifold. Hence, the small perturbations $\delta g_{\mu\nu}$ generate a new geometry $g_{\mu\nu}$ such as
\begin{equation}\label{eq:nashdeformation1}
g_{\mu\nu}=\;\bar{g}_{\mu\nu}+\delta g_{\mu\nu}\;.
\end{equation}
Similarly, the perturbed extrinsic curvature $k_{\mu\nu}$ is given by
\begin{equation}\label{eq:curvextrperturb}
k_{\mu\nu}=\;\bar{k}_{\mu\nu}+\delta k_{\mu\nu}\;,
\end{equation}
where the small perturbations $\delta k_{\mu\nu}=-2\delta y~g^{\sigma\rho}\bar{k}_{\mu\sigma}\bar{k}_{\nu\rho}$. We point out that these geometries must be compatible with the integrability conditions and the present Nash perturbation mechanism makes possible to obtain solutions to the well known Gauss and Codazzi equations~\cite{Greene1970}
\begin{eqnarray}
\nonumber&&^5{\cal R}_{ABCD}\mathcal{Z}^A_{,\alpha}\mathcal{Z}^B_{,\beta}\mathcal{Z}^C_{,\gamma}
\mathcal{Z}^D_{,\delta}= \bar{R}_{\alpha\beta\gamma\delta}\;+\\
&&\hspace{4cm}+\;(\bar{k}_{\alpha\gamma}\;\bar{k}_{\beta\delta}\!-\!\;\bar{k}_{\alpha\delta}\;\bar{k}_{\beta\gamma})\!\;,\label{eq:G1}\\
&&^5{\cal R}_{ABCD}\mathcal{Z}^A_{,\alpha} \mathcal{Z}^B_{,\beta}\mathcal{Z}^C_{,\gamma}\eta^D=\;\bar{k}_{\alpha[\beta;\gamma]} \;, \label{eq:C1}
\end{eqnarray}
where $^5{\cal R}_{ABCD}$ is the five-dimensional Riemann tensor, and $\bar{R}_{\alpha\beta\gamma\delta}$ is the four-dimensional background Riemann tensor. The perturbed embedding coordinate is denoted by $\mathcal{Z}^{A}_{,\mu}$ that is defined as $\mathcal{Z}^{A}_{,\mu}=  \mathcal{X}^{A}_{,\mu} + \delta y \;  \eta^{A}_{,\mu}\;$. We point out that the normal vector $\eta^{A}$ is invariant under perturbations. Then, one obtains the bulk dynamics follow the higher dimensional Einstein equations as
\begin{equation}
^5{\mathcal R}_{AB} -\frac{1}{2} \,{^5{\mathcal R}} {\mathcal G}_{AB}  =G_{*}\; T^*_{AB}\;, \label{eq:BE0}
\end{equation}
where $G_*$ is a new gravitational constant, and $T^*_{AB}$ describes energy-momentum components confined to four dimensions. Matter fields are confined to the embedded space. They remain localized in $V_4$ to ensure no energy leakage into higher dimensions while permitting graviton oscillations~\cite{MAIA20029,GDE}. This works with the definition with the characteristic length $d$ of the extra-dimensional $n$ space accessible to gravitons is given by~\cite{MAIA20029}
\begin{equation}\label{eq:typicallenght}
d=\frac{M_{Pl}^{2/n}}{M_{\ast}^{1+(2/n)}}\frac{1}{(1+\frac{M_{Pl}^2}{M_{e}^2})^{1/n}}   \;,
\end{equation}
where $M_{\ast}$ and $M_{Pl}$ are the fundamental and effective Planck scales, and the extrinsic scale $M_e$ is defined as
\begin{equation}\label{eq:typicallenght2}
\frac{1}{M_{e}}=\int (K^2+h^2)\sqrt{g}d^4x\;.
\end{equation}
This prevents energy leakage to higher dimensions imposing a geometrical constraint, removing the need for a radion field as commonly defined in traditional braneworld models~\cite{Randall1999,Randall1999b,ArkaniHamed1998,Dvali2000}. It is important to point out that the perturbation $\delta y$ in Eq. \eqref{eq:nashdeformation} associated with the embedding direction introduces effective scalar-type perturbations through the deformation of $g_{\mu\nu}$ and $k_{\mu\nu}$, as we see in Eq. \eqref{eq:nashdeformation1} and Eq. \eqref{eq:curvextrperturb}. Thus, the extrinsic curvature perturbations are geometrically induced, , so no negative-norm state arises by construction. In DGP-like models, ghost instabilities arise from the scalar mode associated with the brane bending, when the effective 4D Einstein equations include second derivatives of extrinsic curvature. On the other hand, in our case the extrinsic curvature enters only quadratically and through Gauss-Codazzi constraints. There is no explicit 4D induced curvature term, which such ghostly kinetic terms usually emerge. Then, we avoid the DGP ghost problem.

\section{Main cosmological equations}
Following the derivation proposed in~\cite{MAIA20029,GDE}, the tangent components of Eq. \eqref{eq:BE0} give give the non-perturbed field equations
\begin{eqnarray}
  \bar{G}_{\mu\nu}- \bar{Q}_{\mu\nu} &=&-8\pi G \bar{T}_{\mu\nu}\;,  \label{eq:noperttensoreq}\\
  \bar{k}_{\mu[\nu;\rho]} &=& 0 \;,
  \label{eq:nopervecteq}
\end{eqnarray}
where we use the confinement condition of the energy-momentum  tensor  source $T^*_{AB}$ of the bulk Einstein's equation in Eq. (\ref{eq:BE0})  with the projections
\begin{equation}
8\pi G T_{\mu\nu} =G_* \mathcal{Z}^A_{,\mu}\mathcal{Z}^B_{,\nu}T^*_{AB}, \mathcal{Z}^A_{,\mu}\eta^B T^*_{AB}=0, \eta^A  \eta^B T^*_{AB}=0\;. \label{eq:confinement}
\end{equation}

As a consequence of the embedding, we define the non-perturbed conserved deformation tensor $\bar{Q}_{\mu\nu}$ as
\begin{equation}\label{eq:qmunu}
  \bar{Q}_{\mu\nu}=\bar{k}^{\rho}_{\mu}\bar{k}_{\rho\nu}- \bar{k}_{\mu\nu }h -\frac{1}{2}\left(K^2-h^2\right)\bar{g}_{\mu\nu}\;,
\end{equation}
where $h^2=h\! \cdot \! h$ denotes the mean curvature, being $h= \;\bar{g}^{\mu\nu}\;\bar{k}_{\mu\nu}$, and $K^{2}=\bar{k}^{\mu\nu}\bar{k}_{\mu\nu}$ denotes the Gaussian curvature. The tensor $\bar{Q}_{\mu\nu}$ satisfies $\bar{Q}_{\mu\nu;\mu}=0$.

\subsection{The background equations}
As shown in~\cite{GDE}, applying the Friedmann-Lemaître-Robertson-Walker (FLRW) four-dimensional metric 
\begin{equation}
  ds^2= -dt^2 + a^2\left(dr^2+r^2d\theta^2 + r^2\sin^2\theta d\phi^2 \right)\;,
\end{equation}
one obtains the following quantities
\begin{subequations}\label{eq:BB}
\begin{align}
&\bar{k}_{ij}=\frac{b}{a^2}\bar{g}_{ij},\;\;i,j=1,2,3,\\
&\bar{k}_{44}=\frac{-1}{\dot{a}}\frac{d}{dt}\frac{b}{a},\\
 &
 \bar{k}_{44}=-\frac{b}{a^{2}}\left(\frac{B}{H}-1\right)\;, \\
&K^{2}=\frac{b^2}{a^4}\left( \frac{B^2}{H^2}-2\frac BH+4\right),
  h=\frac{b}{a^2}\left(\frac BH+2\right),\label{eq:hk}\\
&\bar{Q}_{ij}= \frac{b^{2}}{a^{4}}\left( 2\frac{B}{H}-1\right)
\bar{g}_{ij},\;\bar{Q}_{44} = -\frac{3b^{2}}{a^{4}},
  \label{eq:Qab}\\
&\bar{Q}= -(K^2 -h^2) =\frac{6b^{2}}{a^{4}} \frac{B}{H}\;, \label{Q}
\end{align}
\end{subequations}
where $\bar{Q}$ denotes the deformation scalar by the contraction $\bar{g}^{\mu\nu}\bar{Q}_{\mu\nu}=\bar{Q}$. The term $b(t)=b\equiv b(t)=k_{11}=k_{22}=k_{33}$ denotes the bending function as a manifestation of the 
extrinsic effects. Moreover, as a copy of the Hubble parameter $H\equiv H(t)=\frac{\dot{a}}{a}=(da/dt)/a$, we define $B=B(t)\equiv \frac{\dot{b}}{b}=(db/dt)/b$. The $B(t)$ function has the same unit as $H$. The uniqueness of the bending function $b(t)$ is found using the Einstein-Gupta theorem~\cite{Gupta}, which theorem shows  that any  such   tensor  necessarily  satisfy an  Einstein-like  system of  equations. We normalize the extrinsic curvature by $k_{\mu\nu}k^{\mu\nu}  =K^2  \neq  4$ as
\begin{equation}
f_{\mu\nu} = \frac{2}{K}k_{\mu\nu},
\label{eq:fmunu}
\end{equation}
with its inverse by $ f^{\mu\rho}f_{\rho\nu} =   \delta^\mu_\nu$.  Then, one obtains    $f^{\mu\nu}=\frac{2}{k}k^{\mu\nu}$. Denoting  $||$ by the covariant derivative with respect to  a  connection defined by $f_{\mu\nu}$,  while  keeping the usual semicolon notation for the covariant derivative with respect to $g_{\mu\nu}$,  the   analogous to the  ``Levi-Civita"  connection associated with $f_{\mu\nu}$  such   that ''  $f_{\mu\nu||\rho}=0$,  is:
\begin{equation}
\Upsilon_{\mu\nu\sigma}=\;\frac{1}{2}\left(\partial_\mu\; f_{\sigma\nu}+ \partial_\nu\;f_{\sigma\mu} -\partial_\sigma\;f_{\mu\nu}\right)\;.  \label{eq:upsilon}
\end{equation}
Defining $\Upsilon_{\mu\nu}{}^{\lambda}= f^{\lambda\sigma}\Upsilon_{\mu\nu\sigma}$, the  ``Riemann tensor'' for  $f_{\mu\nu}$ has  components
\begin{equation}
{\cal  F}_{\nu\alpha\lambda\mu}= \;\partial_{\alpha}\Upsilon_{\mu\lambda\nu}- \;\partial_{\lambda}\Upsilon_{\mu\alpha\nu}+ \Upsilon_{\alpha\sigma\mu}\Upsilon_{\lambda\nu}^{\sigma} -\Upsilon_{\lambda\sigma\mu}\Upsilon_{\alpha\nu}^{\sigma}
\end{equation}
and  the  corresponding  ``f-Ricci tensor'' and the ``f-Ricci scalar'' for  $f_{\mu\nu}$ are, respectively  by
$
{\cal  F}_{\mu\nu} =  f^{\alpha\lambda}{\cal  F}_{\nu
\alpha\lambda\mu}$
%\;\mathcal{F}_{\mu\lambda\nu}^{\;\;\;\lambda}\;=\mathcal{F}_{\mu\nu}
and ${\cal  F}=f^{\mu\nu}{\cal  F}_{\mu\nu}$. Hence,  Gupta's equations for $f_{\mu\nu}$  can be simply written in a ``Einstein-like vacuum'' form as
\begin{equation}
{\cal  F}_{\mu\nu}=0 \;.  \label{eq:guptaflat}
\end{equation}
Using the spatially flat FLRW metric, we obtain the relations
\begin{equation}
f_{ij}  =\frac{2}{K} g_{ij},\; i,j  = 1..3, \;   f_{44}  = -\frac{2}{K}\frac{1}{\dot{a}}\frac{d}{dt}{\left(\frac{b}{a}\right)}\;.
\label{eq:fij}
\end{equation}
Thus, one solves Eq. (\ref{eq:guptaflat}) obtaining in a generic form~\cite{capistrano2022} for the ratio $B/H$ as
\begin{equation}
\label{eq:EG2}
\frac{B}{H}=\alpha_0\;,
\end{equation}
and by direct integration~\cite{gde2,capistrano2022}, it leads to
\begin{equation}\label{eq:b(t)funct}
b(t)=b_0 a(t)^{\alpha_0}\;,
\end{equation}
where the term $b_0$ is the current value of the bending function and $\alpha_0$ is an integration constant. As a result, the Friedmann equation in terms of redshift can be written as
\begin{eqnarray}\label{eq:nonpertfriedtotal3}
\left(\frac{H}{H_0}\right)^2=\Omega_{m(0)}(1+z)^3+\Omega_{ext(0)}(1+z)^{4-2 \alpha_0},
\end{eqnarray}
where $\Omega_{m(0)}$ denotes the current cosmological parameter for matter density content and the contribution $\Omega_{ext(0)}$ stands for the density parameter associated with the extrinsic curvature, appearing similar \emph{in form only} to a further barotropic term, entering the Friedmann equation \cite{Dunsby:2016lkw}. For a flat universe, one has $\Omega_{ext(0)}= 1- \Omega_{m(0)}$. Moreover, the current extrinsic cosmological parameter $\Omega_{ext(0)}$ is defined as
\begin{equation}\label{eq:extomega}
\Omega_{ext(0)}=\frac{8\pi G}{3H_0^2}\rho_{ext(0)}\equiv \frac{b_0^2}{H_0^2a_0^{\alpha_0}}\;,
\end{equation}
where $a_0$ sets the current value of the scale factor and $\bar{\rho}_{ext}(0)=\frac{3}{8\pi G} b_0^2$ denotes the current extrinsic density.

\subsection{The perturbed cosmic equations}

In the conformal Newtonian gauge, the FLRW metric is given by
\begin{equation}\label{eq:scalarpertmetric2}
ds^2 = a^2 [-(1+ 2\Psi) d\tau^2 + (1-2\Phi) dx^i dx_i] \;,
\end{equation}
where $\Psi=\Psi(\vec{x},\tau)$ and $\Phi=\Phi(\vec{x},\tau)$ denote the Newtonian potential and the Newtonian curvature in conformal time $\tau$ defined as $d\tau=dt/a(t)$. 

The non-perturbed stress energy tensor $\bar{T}_{\mu\nu}$ does not have any modifications due to confinement condition. Then, we define co-moving fluid in the standard form as 
\begin{equation}\label{eq:stresstensor}
\bar{T}_{\mu\nu}=\left(\bar{\rho}+P\right) U_{\mu}U_{\nu} + P \bar{g}_{\mu\nu}\;;\;U_{\mu}=\delta^4_{\mu}\;,
\end{equation}
where $U_{\mu}$ denotes the co-moving velocity. The conservation equation is $
  \frac{d\bar{\rho}}{dt} + 3H\left(\bar{\rho} + P\right)=0$, 
where  $\bar{\rho}$ is the total energy density and $P$ is the total pressure. The  conformal time derivative for the ``contrast'' matter density $\delta_m$ and fluid velocity $\theta$ are written as 
\begin{eqnarray}\label{eq:perturbmatter}
&&\dot{\delta_m} = -(1+w)(\theta-3\dot{\Phi})-3\mathcal{H}(c_s^2-w)\delta_m, \\
&&\dot{\theta} = -\mathcal{H}(1-3w)\theta- \frac{\dot{w}}{1+w}\theta+\frac{c_s^2}{1+w}k^2\delta_m+\nonumber\\
&&\hspace{5cm}-k^2\sigma+k^2\Psi,
\end{eqnarray}
where $\mathcal{H}=aH$ conformal Hubble parameter, $\theta=ik^ju_j$, $w$ is the fluid parameter $w=\frac{P}{\bar{\rho}}$, $c^2_s$ is the sound velocity defined as $c^2_s=\frac{\delta P}{\delta \rho}$ and $\sigma$ is the anisotropic stress. This set of equations in Eq. \eqref{eq:perturbmatter} can be alternatively written as 
\begin{eqnarray}\label{eq:pertmatter}
&&  \delta'_m = 3(1+w)\Phi'- \frac{V}{a^2H}-\frac{3}{a}\left(\frac{\delta P}{\bar{\rho}}-w\delta_m\right)\;, \\
\nonumber&&  V' = -(1-3w)\frac{V}{a}+\frac{k^2}{a^2H}\frac{\delta P}{\bar{\rho}}+(1+w)\frac{k^2}{a^2H}\Psi+\\
&&\hspace{4.5cm}-\frac{k^2}{a^2H}(1+w)\sigma,
\label{eq:velocitypertmatter}
\end{eqnarray}
where $V=(1+w)\theta$ is the scalar velocity. The prime symbol $'$ denotes the ordinary derivative with respect to scale factor as  $'=\frac{d }{da}$. We use this mechanism to avoid divergences when the Equation of State~(EoS) crosses $w=-1$ as pointed out in~\cite{sapone2009,nesseris2019}. 

From Eq. \eqref{eq:nonpertfriedtotal3}, the Friedmann equation can be written in term of densities as
\begin{equation}\label{eq:Friedman2}
H^2=\frac{8}{3}\pi G \left(\bar{\rho}_{m}+\bar{\rho}_{ext}\right)\;\;,
\end{equation}
where $\bar{\rho}_{ext}(a)$ is given by
\begin{equation}\label{eq:extdensitya1}
\bar{\rho}_{ext}(a)=\bar{\rho}_{ext}(0)a^{2 \alpha_0-4}\;,
\end{equation}
with $\bar{\rho}_{ext}(0)=\frac{3}{8\pi G} b_0^2$. Then, the extrinsic ``pressure'' is given by 
\begin{equation}\label{eq:extpressure}
\bar{p}_{ext}(a)=\frac{1}{3}\left(1-2 \alpha_0\right)\bar{\rho}_{ext}(0)a^{2 \alpha_0-4}\;.
\end{equation}
The former results are obtained from the fluid analogy $8\pi G\bar{T}^{ext}_{\mu\nu}\equiv \bar{Q}_{\mu\nu}$ as
\begin{equation}\label{eq:energy tensor2}
-8\pi G\bar{T}^{ext}_{\mu\nu}=(\bar{p}_{ext}+\bar{\rho}_{ext})U_{\mu}U_{\nu}+\bar{p}_{ext}\;\bar{g}_{\mu\nu},\;U_{\mu}=\delta_{\mu}^{4}\;,
\end{equation}
since $T^{ext}_{\mu\nu;\nu}=0$ (as a consequence of the conservation of $\bar{Q}_{\mu\nu}$). Hence, the conservation equation for extrinsic quantities is given by
\begin{equation}\label{eq:pertdensity2}
  \frac{d\bar{\rho}_{ext}}{dt} + 3H\left(\bar{\rho}_{ext} + \bar{p}_{ext}\right)=0\;,
\end{equation}
where $\bar{\rho}_{ext}$ and $\bar{p}_{ext}$ denote the non-perturbed extrinsic ``density'' and extrinsic ``pressure'', respectively. 

The gauge invariant perturbed field equations modified by extrinsic curvature in the Fourier \emph{k}-space wave modes can be written as~\cite{Capistrano2024}
\begin{eqnarray}
&&k^2\Phi_k + 3 \mathcal{H} \left(\Phi^{'}_k + \Psi_k \mathcal{H} \right)= -4\pi G a^2 \delta \rho_k \nonumber\\&&\hspace{4.8cm}+ \beta_0 a^{2 \alpha_0}\Psi_k\;, \label{tensorcompo00kspace}\\
&&\Phi^{'}_{k}+ \mathcal{H}\Psi_k= -4\pi G a^2(\bar{\rho}+P) \frac{\theta}{k^2}\;,\label{tensorcomp0ikspace}\\
&&\mathcal{D}_k + \frac{k^2}{3}(\Phi_k-\Psi_k)= -\frac{4}{3}\pi G a^2 \delta \bar{P}\nonumber\\&&\hspace{4.cm}-\frac{3}{4}\beta_0 a^{2(\alpha_0-1)} \Phi_k\;, \label{tensorcompijkspace}\\
&&k^2(\Phi_{k}-\Psi_k)= 12\pi G a^2(\bar{\rho}+P) \sigma,\label{offdiagonal}
\end{eqnarray}
where $\theta= ik^j\delta u_{\parallel j}$ denotes the divergence of fluid velocity in \emph{k}-space, and $\mathcal{D}_k$ denotes $\mathcal{D}_k=\Phi^{''}_{k} + \mathcal{H}(2\Phi^{'}_k+\Psi^{'}_k) + (\mathcal{H}^2+2\mathcal{H}')\Psi_k$. We have simplified the notation of the aforementioned perturbed equations by merging constants and the extrinsic term $b_0$ of Eq. \eqref{eq:b(t)funct} introducing a dimensionless $\beta_0$ parameter as a relic from extrinsic geometry. It means that $\beta_0\rightarrow 0$, we obtain the GR correspondence limit. 

The perturbed equations given in Eqs. (\ref{tensorcompo00kspace}), (\ref{tensorcomp0ikspace}), (\ref{tensorcompijkspace}), and (\ref{offdiagonal}) can be reorganized into the following set of equations
\begin{eqnarray}
&& k^2\Psi_k = -4\pi a^2 G_{\text{eff}}(a)\rho \Delta\;, \label{modequation1} \\
&&\mathcal{D}_k + \frac{k^2}{3}(\Phi_k-\Psi_k)= -\frac{4}{3}\pi G a^2 \delta \bar{P}\nonumber\\&&\hspace{4.cm}-\frac{3}{4}\beta_0 a^{2(\alpha_0-1)} \Phi_k\;, \label{modtensorcompijkspace}\\
&&k^2(\Phi_{k}-\Psi_k)= 12\pi G a^2(\bar{\rho}+P) \sigma,\label{modoffdiagonal}
\end{eqnarray}
where the density perturbation is given by \(\rho \Delta = \bar{\rho} \delta + 3\frac{\mathcal{H}}{k}(\bar{\rho} + P)\theta\), and the scale independent $G_{\text{eff}(a)}$ is simply
\begin{equation}
G_{\text{eff}}(a) = \frac{G}{1 - \beta_0 a^{2 \alpha_0}} \;. \label{muequation}
\end{equation}
Thus, motivated by Eq. \eqref{eq:energy tensor2}, we define an effective EoS with an ``extrinsic fluid'' parameter $w_{ext}$ by the definition $w_{ext}=\frac{\bar{p}_{ext}}{\bar{\rho}_{ext}}$ to obtain
\begin{equation}\label{eq:wext}
w_{ext}=-1+\frac{1}{3}\left(4-2 \alpha_0\right)\;.
\end{equation}
or alternatively,
\begin{equation}\label{eq:wext2}
\alpha_0=\frac{1}{2}(1-3w), 
\end{equation}
with the fluid correspondence $w_{ext}\doteq w=w_0$. As a result, Eq. \eqref{eq:nonpertfriedtotal3} can be written as the dimensionless Hubble parameter $E(z)=\frac{H(z)}{H_0}$ 
\begin{eqnarray}\label{eq:dimenHub1}
E^2(z)=\Omega_{m(0)}(1+z)^3+\Omega_{ext(0)}(1+z)^{3(1+w_0)}.
\end{eqnarray}
Moreover, $G_{\text{eff}(a)}$ is written accordingly
\begin{eqnarray}
&& G_{\text{eff}}(a) = \frac{G}{1 - \beta_0 a^{1-3w_0}} \;, \label{muequation2}
\end{eqnarray}
similarly to an early-time  scalar field extension or to a non-minimal coupling of the Hilbert-Einstein action \cite{Belfiglio:2022qai,Wolf:2025jed} and implying \emph{de facto} a modification of the gravitational constant.

In what follows, we refer to the scenario proposed here as \textit{Nash Gravity}.

\section{Datasets and methodology}
\label{data}

To validate the theoretical frameworks presented in this study, we integrated our model into the \texttt{CLASS} Boltzmann solver~\cite{Blas:2011rf} and conducted Monte Carlo analyses using the \texttt{MontePython} sampler~\cite{Brinckmann:2018cvx, Audren:2012wb}, leveraging Markov Chain Monte Carlo (MCMC) methods. Convergence was verified via the Gelman-Rubin criterion~\cite{Gelman_1992}, with all runs satisfying the threshold $R - 1 \leq 10^{-2}$.

The cosmological parameters sampled in this analysis include: the physical baryon density, $\omega_{\rm b} = \Omega_{\rm b} h^2$; the physical dark matter density, $\omega_{\rm c} = \Omega_{\rm c} h^2$; the amplitude of the primordial scalar power spectrum, $A_{\mathrm{s}}$; the spectral index of the primordial power spectrum, $n_{\mathrm{s}}$; the optical depth to reionization, $\tau_{\text{reio}}$; and the Hubble constant, $H_0$. In addition to these standard parameters, we also sample the parameter $w_0$, since our background model is $w$CDM, as well as the new parameter $\beta_0$, which is associated with the extrinsic geometry. The adopted prior ranges are: $\omega_{\rm b} \in [0.0, 1.0]$, $\omega_{\rm c} \in [0.0, 1.0]$, $\ln(10^{10} A_{\mathrm{s}}) \in [3.0, 3.18]$, $n_{\mathrm{s}} \in [0.9, 1.1]$, $\tau_{\mathrm{reio}} \in [0.004, 0.125]$, $H_0 \in [20, 100]$, $w_0 \in [-3, -0.5]$, and $\beta_0 \in [\text{None}, \text{None}]$. In all analyses, we used the Python package \texttt{GetDist} \footnote{\url{https://github.com/cmbant/getdist}}, allowing the extraction of numerical results, including 1D posteriors and 2D marginalized probability contours. The datasets employed in this analysis are outlined below.

\begin{itemize}
    \item[1)] \textit{Cosmic Microwave Background} (\textbf{CMB}): We include temperature and polarization data from Planck’s 2018 legacy release~\cite{Planck:2018vyg}, incorporating both auto- and cross-correlations. For high multipoles, we use the \texttt{Plik} likelihood covering TT ($30 \leq \ell \leq 2508$), TE, and EE ($30 \leq \ell \leq 1996$), while low-$\ell$ modes ($2 \leq \ell \leq 29$) are described using the \texttt{SimAll} likelihood~\cite{Planck:2019nip}. We also account for CMB lensing using the 4-point function derived from temperature anisotropies~\cite{Planck:2018lbu}. This ensemble of datasets is collectively denoted as \texttt{CMB}.

    \item[2)] \textit{Baryon Acoustic Oscillations} (\textbf{DESI-DR2}): We use the most recent BAO measurements from DESI’s second data release, which span the redshift range $0.295 \leq z \leq 2.330$ across nine bins. These include constraints from galaxies, quasars~\cite{DESI:2025zgx}, and Lyman-$\alpha$ forest absorption~\cite{DESI:2025zpo}. The distance information is encoded in $D_{\mathrm{M}}/r_{\mathrm{d}}$, $D_{\mathrm{H}}/r_{\mathrm{d}}$, and $D_{\mathrm{V}}/r_{\mathrm{d}}$, where all scales are normalized to the comoving sound horizon at the drag epoch. We also incorporate the correlation coefficients $r_{V,M/H}$ and $r_{M,H}$ to account for statistical dependencies. This dataset is referred to as \texttt{DESI-DR2}.

    \item[3)] \textit{Type Ia Supernovae} (\textbf{SN Ia}): We adopt several recent SN Ia compilations, each providing constraints on luminosity distances:
    \begin{itemize}
        \item \textbf{PantheonPlus} and \textbf{PantheonPlus\&SH0ES} (\texttt{PP}, \texttt{PPS}): The \texttt{PP} dataset~\cite{pantheonplus} includes 1550 SN Ia (1701 light curves) over $0.01 < z < 2.26$, with improvements in calibration and modeling over the original Pantheon. The \texttt{PPS} version uses Cepheid-calibrated anchors from SH0ES~\cite{Riess:2021jrx} to constrain the SN absolute magnitude without assuming an external $H_0$ prior.

        \item \textbf{Union 3.0} (\texttt{Union3}): This compilation~\cite{Rubin:2023ovl} contains 2087 SN Ia events, with 1363 overlapping with PantheonPlus. It employs a hierarchical Bayesian framework to consistently propagate systematics across different datasets.

        \item \textbf{DESY5} (\texttt{DESY5}): As part of the DES Year 5 release~\cite{DES:2024tys}, this dataset comprises 1635 photometrically classified SN Ia in the range $0.1 < z < 1.3$, along with 194 low-$z$ events ($0.025 < z < 0.1$) overlapping with PantheonPlus. The sample benefits from homogeneous selection and well-characterized calibration.
    \end{itemize}
\end{itemize}

\section{Results}
\label{results}

\begin{table*}[htpb!]
\centering\caption{68\% confidence level (CL) constraints for the parameters of the model. In the last rows, we present the quantities $\Delta \chi^2_{\text{min}} \equiv \chi^2_{\text{min (NASH)}} - \chi^2_{\text{min ($\Lambda$CDM)}}$ and $\Delta \text{AIC} \equiv \text{AIC}_{\text{NASH}} - \text{AIC}_{\text{$\Lambda$CDM}}$, which compare the model fits. Negative values for both differences indicate a preference for this model over the $\Lambda$CDM model, while positive values favor the $\Lambda$CDM model.}
\renewcommand{\arraystretch}{1.6}
\resizebox{\textwidth}{!}{
\begin{tabular}{lccccc} \hline
\textbf{Parameter} & \textbf{CMB+DESI-DR2} & \textbf{CMB+DESI-DR2+PP} & \textbf{CMB+DESI-DR2+PPS} & \textbf{CMB+DESI-DR2+Union3} & \textbf{CMB+DESI-DR2+DESY5}\\ \hline \hline
$10^{2} \Omega_{\rm b} h^2$ & $2.254\pm 0.013$ ($2.254^{+0.026}_{-0.028}$) & $2.259\pm 0.012$ ($2.259^{+0.023}_{-0.024}$) & $2.260\pm 0.013$ ($2.260^{+0.024}_{-0.025}$) & $2.259\pm 0.013$ ($2.259^{+0.026}_{-0.025}$) & $2.263\pm 0.013$ ($2.263^{+0.025}_{-0.025}$)  \\
$\Omega_{\rm c} h^2$ & $0.11781\pm 0.00065$ ($0.1178^{+0.0013}_{-0.0012}$) & $0.11721\pm 0.00059$ ($0.1172^{+0.0012}_{-0.0011}$) & $0.11766\pm 0.00059$ ($0.1177^{+0.0012}_{-0.0011}$) & $0.11724\pm 0.00064$ ($0.1172^{+0.0013}_{-0.0013}$) & $0.11673\pm 0.00061$ ($0.1167^{+0.0012}_{-0.0012}$)  \\
$\log(10^{10} A_\mathrm{s})$ & $3.045^{+0.015}_{-0.016}$ ($3.045^{+0.035}_{-0.031}$) & $3.045\pm 0.015$ ($3.045^{+0.029}_{-0.029}$) & $3.043\pm 0.016$ ($3.043^{+0.033}_{-0.032}$) & $3.044\pm 0.016$ ($3.044^{+0.032}_{-0.029}$) & $3.047^{+0.015}_{-0.017}$ ($3.047^{+0.034}_{-0.031}$)  \\
$n_{s }$ & $0.9711\pm 0.0034$ ($0.9711^{+0.0064}_{-0.0067}$) & $0.9731\pm 0.0033$ ($0.9731^{+0.0067}_{-0.0067}$) & $0.9720\pm 0.0033$ ($0.9720^{+0.0067}_{-0.0065}$) & $0.9731\pm 0.0033$ ($0.9731^{+0.0065}_{-0.0063}$) & $0.9744\pm 0.0035$ ($0.9744^{+0.0067}_{-0.0069}$)  \\
$\tau{}_{reio }$ & $0.0567\pm 0.0079$ ($0.057^{+0.016}_{-0.015}$) & $0.0578\pm 0.0073$ ($0.058^{+0.014}_{-0.014}$) & $0.0559\pm 0.0079$ ($0.056^{+0.016}_{-0.016}$) & $0.0569\pm 0.0077$ ($0.057^{+0.016}_{-0.015}$) & $0.0588^{+0.0074}_{-0.0084}$ ($0.059^{+0.017}_{-0.015}$)  \\
$\Omega{}_{m }$ & $0.2935\pm 0.0055$ ($0.294^{+0.011}_{-0.010}$) & $0.3017\pm 0.0042$ ($0.3017^{+0.0083}_{-0.0083}$) & $0.2919\pm 0.0038$ ($0.2919^{+0.0073}_{-0.0072}$) & $0.3008\pm 0.0047$ ($0.3008^{+0.0094}_{-0.0092}$) & $0.3074\pm 0.0039$ ($0.3074^{+0.0077}_{-0.0079}$)  \\
$S_8$ & $0.779^{+0.034}_{-0.028}$ ($0.779^{+0.058}_{-0.061}$) & $0.767\pm 0.027$ ($0.767^{+0.051}_{-0.052}$) & $0.768\pm 0.030$ ($0.768^{+0.059}_{-0.059}$) & $0.764\pm 0.029$ ($0.764^{+0.055}_{-0.061}$) & $0.759\pm 0.029$ ($0.759^{+0.057}_{-0.059}$)  \\
$H_0 [\text{km}/\text{s}/\text{Mpc}]$ & $69.32\pm 0.72$ ($69.3^{+1.4}_{-1.4}$) & $68.24\pm 0.50$ ($68.2^{+1.0}_{-0.99}$) & $69.48\pm 0.48$ ($69.48^{+0.92}_{-0.92}$) & $68.34\pm 0.59$ ($68.3^{+1.2}_{-1.2}$) & $67.49\pm 0.46$ ($67.49^{+0.91}_{-0.88}$)  \\
$\omega_0$ & $-1.030\pm 0.029$ ($-1.030^{+0.056}_{-0.057}$) & $-0.986\pm 0.020$ ($-0.986^{+0.039}_{-0.040}$) & $-1.031\pm 0.019$ ($-1.031^{+0.037}_{-0.037}$) & $-0.990\pm 0.024$ ($-0.990^{+0.046}_{-0.048}$) & $-0.956\pm 0.019$ ($-0.956^{+0.037}_{-0.037}$)  \\
$\beta{}_{0 }$ & $0.22^{+0.24}_{-0.20}$ ($0.22^{+0.39}_{-0.42}$) & $0.29^{+0.20}_{-0.14}$ ($0.29^{+0.30}_{-0.35}$) & $0.28^{+0.22}_{-0.16}$ ($0.28^{+0.37}_{-0.40}$) & $0.30^{+0.20}_{-0.15}$ ($0.30^{+0.34}_{-0.36}$) & $0.32^{+0.19}_{-0.15}$ ($0.32^{+0.32}_{-0.35}$)  \\
\hline
$\Delta \chi^{2}_{\mathrm{min}}$ & -2.34 & -2.12 & -4.22 & -2.34 & -6.20 \\
$\Delta \mathrm{AIC}$ & -0.34 & -0.12 & -2.22 & -0.34 & -4.20 \\
\hline \hline \end{tabular}}
\label{tab:resultsnew}
\end{table*}

\begin{figure*}[htpb!]
    \centering
    \includegraphics[width=0.8\textwidth]{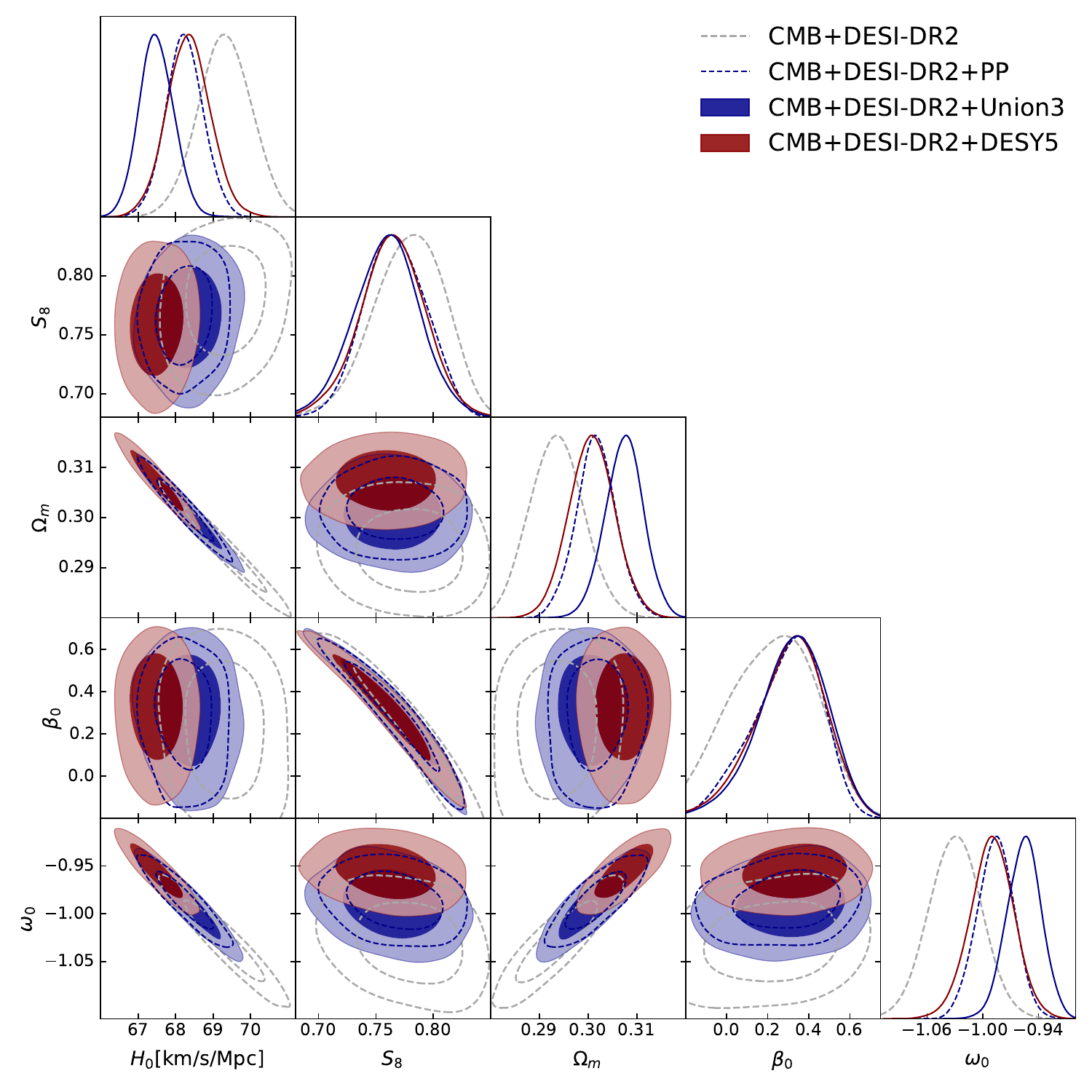} \,\,\,
    \caption{Marginalized posterior distributions and 68\% and 95\% CL contours for the parameters in the model. The results are shown for different combinations of datasets, as indicated in the legend.}
    \label{fig:constraintsnew}
\end{figure*}

Table~\ref{tab:resultsnew} presents a comprehensive summary of the observational constraints on all baseline cosmological parameters of the model considered in this work, as obtained from the full set of analyses performed.

We begin by focusing on the joint analysis of the \textit{CMB} and \textit{DESI-DR2} datasets. This combination provides stringent constraints on the key parameters governing deviations from standard background expansion and linear perturbation theory. Specifically, we obtain:
\[
w_0 = -1.030 \pm 0.029, \quad \text{and} \quad \beta_0 = 0.22^{+0.24}_{-0.20}.
\]
These results suggest a mild departure from the standard $\Lambda$CDM scenario at the $68\%$ confidence level. However, they remain fully consistent with GR and the $\Lambda$CDM model within $95\%$ confidence, as also reported in Table~\ref{tab:resultsnew}. The value of \( w_0 < -1 \) hints at a possible phantom-like behavior, although still statistically compatible with \( w = -1 \) within current uncertainties.

An important feature of the results is the prediction of a slightly lower value for the parameter \( S_8 \), which quantifies the amplitude of matter fluctuations and is known to be in tension between large-scale structure and CMB observations in the standard cosmological model (see \cite{CosmoVerse:2025txj}). From our analysis, we find:
\[
S_8 = 0.779^{+0.034}_{-0.028},
\]
a value that lies below the $\Lambda$CDM prediction. The reduction in \( S_8 \) can be attributed to modifications in the effective gravitational coupling induced by the parameter \( \beta_0 \), which directly impacts the growth of cosmic structures. Specifically, as shown in equation~(\ref{muequation}), the effective gravitational constant increases for positive values of \( \beta_0 \), leading to a suppression of matter perturbation growth. 

This effect is further illustrated in Figure~\ref{fig:constraintsnew}, which displays the two-dimensional marginalized posterior distributions (68\% and 95\% confidence contours) in the parameter spaces involving \( w_0 \), \( \beta_0 \), \( H_0 \), \( S_8 \), and \( \Omega_m \). Notably, we observe a strong negative correlation between \( S_8 \) and \( \beta_0 \), particularly for \( \beta_0 > 0 \), which directly reflects the physical mechanism mentioned above. Thus, the observational constraints obtained in this analysis indicate that the model under consideration is both consistent with current cosmological data and capable to predict lower observed values of \( S_8 \). 

All other cosmological parameters exhibit values that are consistent with the predictions of the $\Lambda$CDM model within current observational uncertainties. However, a noteworthy feature of our results is the inferred value of the Hubble constant, $H_0 = 69.32 \pm 0.72$ km/s/Mpc, which is slightly higher than the standard $\Lambda$CDM prediction of $H_0 = 68.17 \pm 0.28$ km/s/Mpc. This shift is primarily driven by the strong degeneracy between $w_0$ and $H_0$, as shown in Figure~\ref{fig:constraintsnew}, where a less negative EoS parameter $w_0$ allows for an increased expansion rate today. Although this result does not fully resolve the current $H_0$ tension between early- and late-time measurements, it provides a modest alleviation of the discrepancy. More importantly, it highlights that the background dynamics within the Nash gravity framework can introduce modified correlations among cosmological parameters, particularly those involving $w_0$, $H_0$, and $\Omega_m$, which are absent in standard $\Lambda$CDM. These new correlations offer a physically motivated mechanism for accommodating slightly larger values of $H_0$ without violating other observational constraints, thereby contributing to the ongoing effort to understand the origin of the Hubble tension.

In what follows, we interpret the impact of including different SN Ia samples in our analysis. Currently, three major SN Ia compilations are available: PP, Union 3.0, and DES-Y5. In Table~\ref{tab:resultsnew}, we summarize the results obtained by combining each of these samples with CMB and BAO data from DESI-DR2. The inclusion of SN Ia data systematically improves the constraints on the cosmological parameters by breaking degeneracies present in the CMB+BAO combination alone.

Focusing on the background expansion, we find the following constraints on the EoS parameter:
\begin{align*}
w_0 &= -0.986 \pm 0.020 \quad (\text{CMB+DESI+PP}) \\
w_0 &= -1.031 \pm 0.019 \quad (\text{CMB+DESI+PPS}) \\
w_0 &= -0.990 \pm 0.024 \quad (\text{CMB+DESI+Union3.0}) \\
w_0 &= -0.956 \pm 0.019 \quad (\text{CMB+DESI+DESY5})\;.
\end{align*}
Notably, the combinations with PPS and DES-Y5 data show statistically significant deviations from the $\Lambda$CDM value ($w_0 = -1$), with more than $2\sigma$ significance. This suggests that the choice of SN Ia dataset plays a non-negligible role in constraining deviations from the standard cosmological model at the background level.

At the level of scalar perturbations, the joint analysis that includes DESY5 is the only one predicting a deviation in the coupling parameter $\beta_0$ at a comparable level of statistical significance. This reinforces the idea that specific SN Ia samples, particularly DESY5, can impact not only the background expansion history but also the growth of structures when interpreted in the context of Nash Gravity (see $\beta_0$ values in Table~\ref{tab:resultsnew}).

Turning to the $S_8$ parameter, we observe the following constraints:
\begin{align*}
S_8 &= 0.767 \pm 0.027 \quad (\text{CMB+DESI+PP}) \\
S_8 &= 0.768 \pm 0.030 \quad (\text{CMB+DESI+PPS}) \\
S_8 &= 0.764 \pm 0.029 \quad (\text{CMB+DESI+Union3.0}) \\
S_8 &= 0.759 \pm 0.029 \quad (\text{CMB+DESI+DESY5})\;.
\end{align*}
These results consistently show a tendency for $S_8$ to remain at relatively low values compared to those predicted by $\Lambda$CDM. The error bars are also notably reduced due to the improved constraints on $\Omega_m$ provided by the inclusion of SN Ia data.

Thus, we can see that the Nash Gravity framework not only accommodates deviations in the EoS, but also provides a physically motivated mechanism for lowering the amplitude of matter fluctuations, thus addressing a possible $S_8$ tension. This robustness across different SN Ia datasets reinforces the model's ability to capture essential features of the late-time Universe in a statistically and physically consistent manner.

It is important to mention that recent weak lensing measurements, particularly from KiDS-1000 \cite{Wright:2025xka}, are now largely consistent with Planck-CMB predictions. In contrast, full-shape galaxy clustering studies still reveal a notable tension, reaching up to 4.5$\sigma$ \cite{Chen:2024vuf,Ivanov:2024xgb}, compared to early-universe expectations. Additional indications of a discrepancy in $S_8$ appear in RSD measurements \cite{Nunes:2021ipq,Kazantzidis:2018rnb} and other late-time probes \cite{Dalal:2023olq,Karim:2024luk}. Consequently, low $S_8$ values may reflect either a genuine cosmological tension or unaccounted-for systematic effects. In this context, Nash Gravity offers the possibility of naturally producing lower $S_8$ values, assuming that the observed differences are not entirely driven by systematics.

\begin{figure*}[htpb!]
    \centering
    \includegraphics[width=0.47\textwidth]{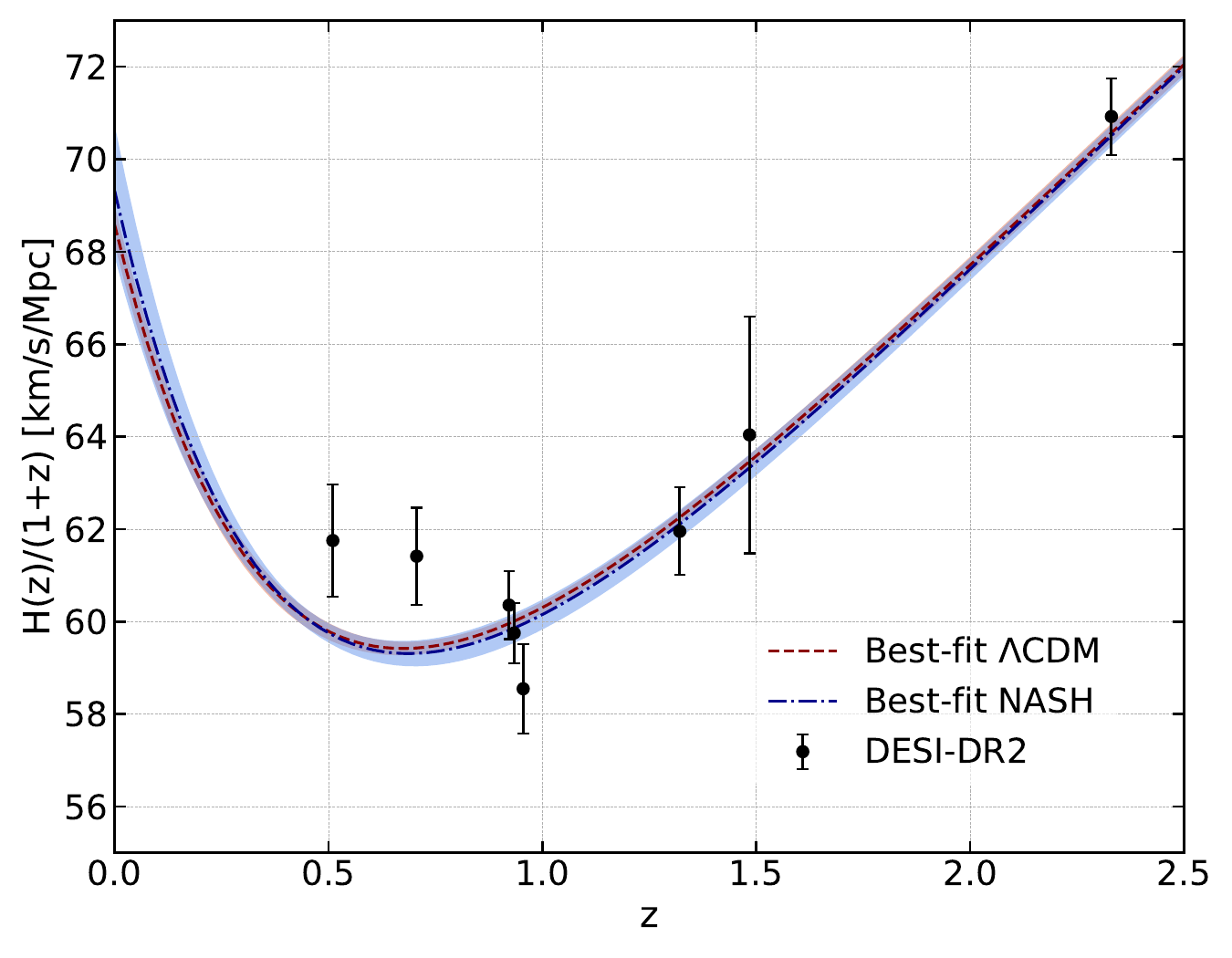} \,\,\,
    \includegraphics[width=0.48\textwidth]{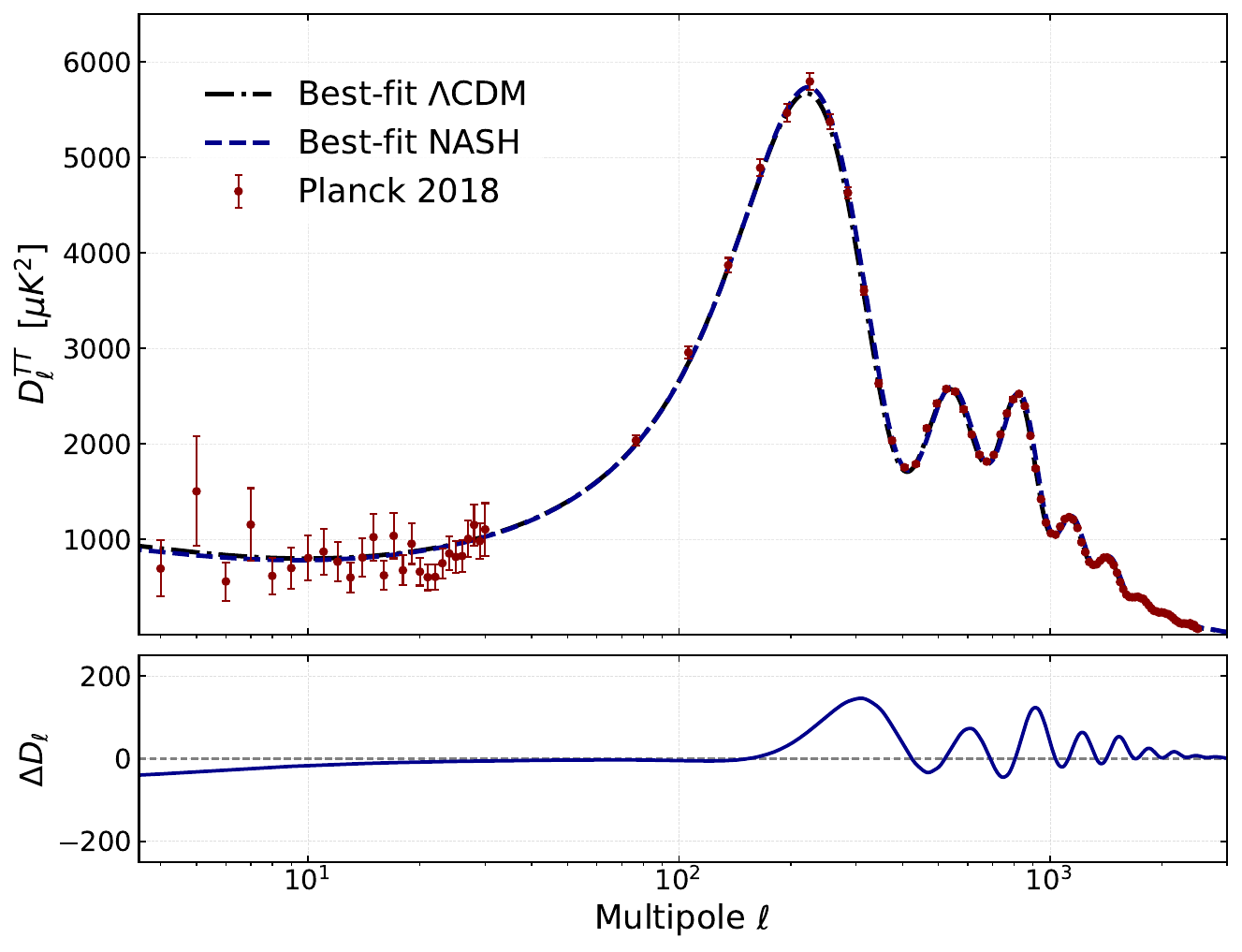} 
    \caption{\textbf{Left panel:} Statistical reconstruction of the (rescaled) expansion rate of the universe, $H(z)/(1+z)$, at $2\sigma$ confidence levels for the $\Lambda$CDM and NASH models, based on the joint analysis of CMB+DESI-DR2, compared to DESI-DR2 measurements. \textbf{Right panel:} Comparison between the theoretical predictions of the $\Lambda$CDM and NASH models for the temperature anisotropy power spectrum. All cosmological parameters are fixed to their respective best-fit values obtained from the joint analysis of CMB + DESI-DR2 + DESY5 data. The error bars on the data points indicate $\pm 1\sigma$ uncertainties. The lower panel displays the relative deviation between the NASH and $\Lambda$CDM predictions.}
    \label{fig:H(z)}
\end{figure*}

To gain further insight into the compatibility (or discrepancy) between the models under investigation and the observational datasets employed, we carry out a statistical comparison with the reference $\Lambda$CDM model by applying the well-established Akaike Information Criterion (AIC)~\cite{akaike1974}. The AIC is expressed as:
\begin{equation}
\mathrm{AIC} = -2 \ln \mathcal{L}_{\mathrm{max}} + 2N,
\end{equation}
where $\mathcal{L}_{\mathrm{max}}$ denotes the maximum likelihood achieved by the model, and $N$ represents the total number of independent parameters.

This criterion serves as a tool for model selection by weighing the trade-off between the quality of the fit and the complexity of the model. A lower AIC score implies a more statistically preferred model, with the formula incorporating a penalty for the inclusion of extra parameters that do not lead to a significant improvement in the fit. In this way, the AIC guards against overfitting and facilitates a meaningful comparison between models of varying degrees of freedom.

Table~\ref{tab:resultsnew} presents the differences in the minimum chi-squared and Akaike Information Criterion values between Nash Gravity and the standard $\Lambda$CDM model, defined respectively as $\Delta \chi^2_{\text{min}} \equiv \chi^2_{\text{min (Nash Gravity)}} - \chi^2_{\text{min} (\Lambda\text{CDM})}$ and $\Delta \text{AIC} \equiv \text{AIC}_{\text{Nash Gravity}} - \text{AIC}_{\Lambda\text{CDM}}$. These quantities allow for a direct statistical comparison of the models' performance in fitting the observational data.

Following standard interpretation~\cite{akaike1974, kass1995}, a value of $\Delta \text{AIC} > 5$ is generally regarded as strong evidence in favor of the model under test—in this case, Nash Gravity-relative to the reference model. Based on this criterion, we find that the joint analysis using CMB + DESI-DR2 + DESY5 data exhibits a significant statistical preference for Nash Gravity over the GR-based $\Lambda$CDM scenario. Notably, this is the only main joint analysis in which such a preference is observed. This result highlights the potential of Nash Gravity to provide a viable alternative explanation for late-time cosmic acceleration while remaining consistent with current large-scale structure and background observations.
However, it is important to note that commonly used parametrizations, such as the CPL model and other theoretical frameworks, can achieve an even better fit, with $\Delta \chi^2_{\text{min}} \approx -20$ to $-15$ (e.g., \cite{DESI:2025zgx,Scherer:2025esj,Wolf:2025acj}), for joint analyses like CMB + DESI-DR2 + DESY5. Therefore, while Nash Gravity can outperform the standard $\Lambda$CDM in certain joint analyses, its overall performance remains inferior to that of dynamical models such as CPL and other scenarios recently explored in the literature.

Although some of our analyses reveal notable deviations in the parameters $w_0$ and $\beta_0$, particularly in joint constraints involving specific SN Ia samples, these deviations must be interpreted cautiously. When accounting for the introduction of additional degrees of freedom and applying the appropriate statistical penalties-such as those implemented by information criteria like the AIC- the extended models are not statistically favored over the standard $\Lambda$CDM scenario with strong significance. In other words, despite localized shifts in parameter values, the $\Lambda$CDM model remains broadly consistent with the data and cannot be decisively ruled out by the present analysis.

Figure \ref{fig:H(z)} shows on the left panel the reconstruction of the $H(z)$ function obtained from the joint CMB+DESI-DR2 analysis. It compares the best-fit curves, along with their 2$\sigma$ confidence regions, for the Nash Gravity and $\Lambda$CDM models against the some observational DESI data. Within $2\sigma$, the results from this joint analysis indicate that the models are statistically indistinguishable—that is, without the inclusion of SNIa data, Nash Gravity and the standard $\Lambda$CDM model yield equivalent fits to the background and large-scale structure observables. 
Nevertheless, it is noteworthy that despite their statistical equivalence, the Nash Gravity scenario systematically predicts a higher Hubble expansion rate at low redshifts, i.e., $H(z)_{\text{Nash}} > H(z)_{\Lambda\text{CDM}}$, when evaluated along the best-fit curve. This directly reflects the point discussed earlier: the background dynamics of Nash Gravity naturally allow for larger values of the Hubble constant $H_0$ compared to the standard cosmological model. This feature may offer a partial resolution or alleviation of the $H_0$ tension observed between early- and late-universe measurements.

Figure~\ref{fig:H(z)} presents, in the right panel, a comparison between the theoretical predictions of the $\Lambda$CDM model (red curve) and the Nash Gravity scenario (blue curve) for the CMB temperature anisotropy power spectrum. All cosmological parameters have been fixed to their respective best-fit values obtained from the CMB+DESI-DR2 joint analysis. The lower panel shows the relative difference between the two models across multipole scales.
We observe that at large angular scales (low multipoles), there is a small but noticeable deviation between the models. This difference is primarily attributed to modifications in the evolution of the gravitational potential, as governed by the perturbed Einstein equations in the Nash Gravity framework (see Eqs.~\ref{tensorcompo00kspace}- \ref{tensorcompijkspace}). These perturbative effects alter the late-time Integrated Sachs-Wolfe effect, which is particularly sensitive to the dynamics of cosmic acceleration and time variation of the gravitational potential.
Additionally, deviations in the background expansion history- especially those driven by the value of $w_0$- also contribute to modifications at large angular scales. Consequently, Nash Gravity predicts distinct imprints on the late-time ISW signal, arising from both background and perturbative departures from standard GR-based cosmology. At smaller angular scales (high multipoles), the observed differences are more subtle and may result from perturbed Einstein equations can change known aspects on these scales. These features highlight how Nash Gravity can introduce modifications in the CMB power spectrum, offering testable predictions distinct from $\Lambda$CDM.

\section{Final Remarks}
\label{conclu}

In this work, we have investigated the observational viability of Nash Gravity as an alternative framework to the standard $\Lambda$CDM cosmology. Unlike traditional braneworld models that rely on auxiliary scalar fields or fine-tuned brane tensions, Nash Gravity introduces orthogonal perturbations of the embedded geometry by means of dynamic variations of the extrinsic curvature. These deformations generate scalar-type metric perturbations directly from geometry, with no need for additional degrees of freedom and avoiding ghost instabilities characteristic of DGP-like models. The embedding framework is constrained by the Gauss-Codazzi equations and preserves the integrability conditions, ensuring a consistent evolution of both intrinsic and extrinsic geometries. Matter fields are confined to the 4D hypersurface, while gravitational degrees of freedom propagate into the bulk. A natural length scale emerges from the extrinsic curvature, setting a geometric bound on gravitational leakage. 
%\abe{The present work revises and simplifies this framework, unifying background and perturbative dynamics adopting a transparent, data-driven approach to late-time cosmology. We extend the analysis of previous papers [71-74] with new theoretical advances are summarized as follows:}

%\begin{itemize}
%\item \abe{Reformulation of the 5D embedding dynamics directly in terms of the extrinsic curvature scalar $k_{\mu\nu}k^{\mu\nu}$ and its dependence on the scale factor $a(t)$, yielding a simpler and more transparent dynamical scheme linking geometry and cosmology;}
%\item \abe{Definition of a new dimensionless coupling $\beta_0$ that captures the residual influence of extrinsic curvature perturbations, clarifying the deviations from GR growth dynamics;}
%\item \abe{Derivation of the full set of linear perturbation equations for Nash gravity (Eqs. 39–46), incorporating the coupled evolution of the $\Phi$ and $\Psi$ potentials, a geometric modification of the Poisson equation via the $\beta_0$ parameter, and the resulting scale-independent effective gravitational constant $G_\text{eff}(a)$};
%\item \abe{Perturbations remain ghost-free, because the extrinsic curvature enters quadratically and satisfies Codazzi conditions, explicitly verified using the Gauss–Codazzi constraints.}
%\end{itemize}

Using recent data from the CMB, DESI DR2 BAO, and several SNIa samples, we performed a comprehensive parameter estimation analysis, evaluated scalar perturbations, and compared predictions with those from GR. Our results show that Nash Gravity provides a consistent fit to current observational data. In particular, we find that Nash Gravity predicts slightly higher values of the Hubble constant $H_0$, with $H_0 = 69.32 \pm 0.72$ km/s/Mpc, compared to the $\Lambda$CDM prediction of $68.17 \pm 0.28$ km/s/Mpc. Although this does not fully resolve the well-known $H_0$ tension, it represents a meaningful alleviation within the framework of a modified background evolution.

At the level of scalar perturbations, we observe a consistent prediction of lower \( S_8 \) values across all SN Ia joint analyses, with $S_8 \approx 0.76$, suggesting that Nash gravity naturally suppresses structure formation. This trend is further supported by the behavior of the effective gravitational coupling, $G_{\rm eff}$, which induces a strong negative correlation with $S_8$ through a well-motivated modification of General Relativity. Notably, the joint analysis including CMB+DESI-DR2+DESY5 exhibits $\Delta\text{AIC} > 5$, providing strong statistical support for Nash Gravity as a viable cosmological model. Our findings open new avenues for testing gravitational dynamics beyond GR. Nash Gravity emerges as a compelling alternative that remains consistent with current observations while having the potential to address key cosmological tensions, particularly the $H_0$ and $S_8$ discrepancies-with special emphasis on the latter.

Given the strong potential of the scenario to predict low values of $S_8$, future work will focus on further exploring the model's predictions by directly confronting data on the formation and evolution of large-scale structures. In particular, we plan to investigate possible scale-dependent effects on the effective gravitational coupling function, $G_{\rm eff}(k, z)$, and their imprints on current observational datasets. This approach will enable us to refine constraints on the parameter $\beta_0$ and better assess the long-term viability of Nash Gravity as a theoretical framework. In summary, Nash Gravity offers a physically motivated, observationally consistent, and statistically competitive extension of the standard cosmological model, justifying ongoing and future studies in the context of late-time cosmic acceleration. In particular, allowing $\alpha$ to vary with the scale factor may yield a dynamical $w_{\text{ext}}(a)$ as an extension of the present work. However, it requires a more careful treatment of the extrinsic curvature evolution equation and its feedback on perturbations, introducing additional functional degrees of freedom that will be explored in future developments.

\begin{acknowledgments}
We thank the referee for their thoughtful comments, which have helped to improve both the clarity and the overall impact of our results. A.J.S.C acknowledges Conselho Nacional de Desenvolvimento Científico e Tecnologico (CNPq, National Council for Scientific and Technological Development) for the partial financial support for this work (Grant No. 305881/2022-1) and Fundação da Universidade Federal do Paraná (FUNPAR, Paraná Federal University Foundation) through public notice 04/2023-Pesquisa/PRPPG/UFPR for the partial financial support (Process No. 23075.019406/2023-92), and the financial support of the NAPI ``Fenômenos Extremos do Universo" of Fundação de Apoio à Ciência, Tecnologia e Inovação do Paraná, (NAPI FÍSICA – FASE 2), under protocol No 22.687.035-0. E.S. received support from the CAPES (Coordination for the Improvement of Higher Education Personnel) scholarship. R.C.N. thanks the financial support from the CNPq under the project No. 304306/2022-3, and the Fundação de Amparo à Pesquisa do Estado do RS (FAPERGS, Research Support Foundation of the State of RS) for partial financial support under the project No. 23/2551-0000848-3. O.L. acknowledges support by the  Fondazione  ICSC, Spoke 3 Astrophysics and Cosmos Observations National Recovery and Resilience Plan (Piano Nazionale di Ripresa e Resilienza, PNRR) Project ID $CN00000013$ ``Italian Research Center on  High-Performance Computing, Big Data and Quantum Computing" funded by MUR Missione 4 Componente 2 Investimento 1.4: Potenziamento strutture di ricerca e creazione di ``campioni nazionali di R\&S (M4C2-19 )" - Next Generation EU (NGEU).
\end{acknowledgments}

%\appendix

%\section{Appendixes}


\end{document}